# Electronic structure of GaAs$_{1-x}$N$_x$ alloy by soft-X-ray absorption and emission: Origin of the reduced optical efficiency

V.N. Strocov[1], P.O. Nilsson[2], A. Augustsson[3], T. Schmitt[3], D. Debowska-Nilsson[2], R. Claessen[1], A.Yu. Egorov[4], V.M. Ustinov[4], Zh.I. Alferov[4]

[1] *Experimentalphysik II, Universität Augsburg, D-86135 Augsburg, Germany (email F1XVS@fy.chalmers.se),* [2] *Department of Physics, Chalmers University of Technology, SE-412 96 Göteborg, Sweden,* [3] *Department of Physics, Uppsala University, Box 530, S-75121 Uppsala, Sweden,* [4] *A.F. Ioffe Physico-Technical Institute, 194021 St. Petersburg, Russia*

(Submitted July 10, 2002)

**Introduction** GaAs$_{1-x}$N$_x$ and related semiconductor alloys are new promising materials, whose potential applications range from efficient solar cells to long wavelength ($\lambda \sim 1.3 \mu m$) optoelectronics, including laser diodes. These alloys show an extremely strong narrowing of the band gap upon increase of the N mole fraction, because the N atoms inserted into GaAs yield a giant perturbation of the electronic structure, forming N impurity and cluster states on top of perturbed host states (see a review in [1]). However, this is accompanied by a significant reduction of the optical efficiency whose origin is still unclear.

Existing experimental data on the electronic structure of GaAs$_{1-x}$N$_x$ are mainly due to various optical spectroscopies such as photoluminescence (PL) (for a compilation see [1]). However, they in general give only the position of the energy levels. An information on spatial localization of the wavefunctions and their orbital character can be achieved by soft-X-ray emission (SXE) and absorption (SXA) spectroscopies, for the valence band (VB) and conduction band (CB) respectively (see, e.g., [2]). As the selection rules are determined here by the core state, they give a view of the electronic structure complementary to that by the optical spectroscopies. Because of small atomic concentrations in diluted alloys such as GaAs$_{1-x}$N$_x$ and small crossection of the SXA and SXE processes, use of 3-rd generation synchrotron radiation sources providing soft-X-rays at high flux and brilliance is required.

We here present the first SXE/SXA data on the N-local electronic structure of GaAs$_{1-x}$N$_x$. This is used, in particular, to identify the origin of the reduced optical efficiency.

**Experiment** The GaAs$_{1-x}$N$_x$ samples were grown by molecular beam epitaxy (MBE) at an EP-1203 machine (Russia) equipped by a solid-phase As source and a radio-frequency plasma N source. The growth was performed on a GaAs (001) substrate at 430°C under As-rich conditions. The GaAs$_{1-x}$N$_x$ layer had a thickness of 2000 Å and a N concentration of $x$=3% (as determined from by X-ray diffraction rocking curves). A 150 Å thick GaAs/AlAs/GaAs cap layer was grown on top. A high temperature annealing was performed after deposition of the AlAs layer, lasting 10-20 minutes at 700-750°C. This typically increases the PL intensity by a factor of 10-20 [3], and simultaneously improves the homogeneity of the N concentration to reduce electronic structure fluctuations [4].

The SXE/SXA experiments were performed at the MAX-lab, Sweden, using the undulator beamline I511-3 of the MAX-II storage ring. The beamline is equipped with a plane grating monochromator and a high-resolution Rowland-mount grazing incidence spectrometer [5]. The measurements used the N 1$s$ core level at ~400 eV. The SXA spectra were recorded in the total fluorescence yield to penetrate through the thick cap layer. The monochromator resolution was ~0.12 eV FWHM, comparable with the N 1$s$ lifetime width. The synchrotron radiation excited SXE spectra were recorded with a monochromator resolution of ~1.5 eV (0.5 eV for the resonance spectra). The SXE spectrometer was operated at a resolution of ~1.4 eV. It was calibrated in the monochromator energy scale using the elastic peaks in the SXE spectra, which enables direct comparison of the VB represented by SXE and CB represented by SXA (absolute photon energies are not important here).

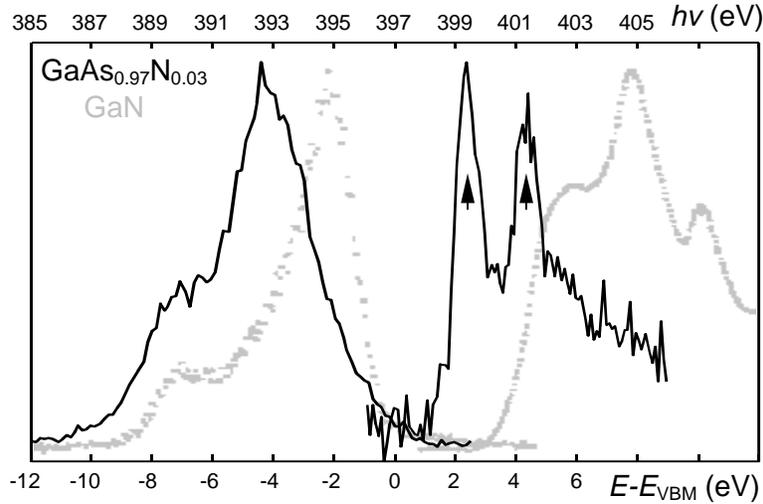

Fig. 1. Experimental N1$s$ SXA and off-resonant SXE spectra of GaAs$_{0.97}$N$_{0.03}$, reflecting the N-local $p$-DOS. The SXA peaks used in the resonance SXE measurements are marked by arrows. The GaN spectra [6], aligned by the VB leading edge, are shown by watermarks.

**N-local electronic structure** Our experimental N1$s$ SXA spectrum and off-resonant SXE spectrum (excitation energy of 420 eV, well above the absorption threshold) are shown in Fig.1. They reflect, by the dipole selection rules requiring that the orbital quantum number $l$ changes by ±1, the $p$-component (predominating in the whole VB and CB) of the local DOS in the N core region. Core exciton effects such as a recombination peak in resonant SXE spectra were not seen. The strong SXA signal near the CB minimum (CBM) reflects involvement of N states in narrowing of the band gap. Among polymorphic local environments of the N atoms in GaAs$_{1-x}$N$_x$ [1], the SXE/SXA spectra reflect mainly the isolated N impurities, because with $x$=3% the concentration of clusters, assuming random statistics, is only ~11.5% of that of the impurities.

Our results can be compared with the N 1$s$ spectra measured on wurtzite GaN [7], also shown in Fig.1 (the spectra of GaN in the zinc-blende metastable structure [6] are similar). In the VB of GaAs$_{0.97}$N$_{0.03}$, the SXE leading edge at the VB maximum (VBM) is much less steep than that of GaN, and the low-energy shoulder at the VB bottom is reduced. In the CB, the two dominant SXA peaks are entirely new to the GaN absorption signal, and the two higher energy structures are suppressed. These differences manifest a radical change of the local electronic structure of the N atoms in the GaAs$_{1-x}$N$_x$ alloy compared to crystalline GaN.

Now we are in position to trace the origin of the reduced optical efficiency of GaAs$_{1-x}$N$_x$ alloys. Insertion of N into GaAs results in localization of the CBM wavefunction on the N atoms. Simultaneously, the VBM charge is displaced from the N atoms, as evidenced by strong suppression of the SXE signal at the VB leading edge. Such a shift of the CBM and VBM wavefunctions in opposite directions results in their weak overlap, reducing the optical efficiency. Interestingly, replacement of a fraction of Ga atoms by In recovers much of the efficiency [3,4].

A detailed interpretation of SXE/SXA requires knowledge of the dipole transition matrix elements. Tentatively, we can reconcile our data with available calculations [1] based on the spatial localization of the CB states. The CB states with maximal N localization are the $t_2(L_{1c})$ perturbed host state and $a_1(N)$ resonant impurity state (the $a_1(\Gamma_{1c})$ derived $E_-$ state and $a_1(L_{1c})$ derived $E_+$ states are more delocalized). We assign them to the two dominant SXA peaks. Their energies are consistent with the theoretical concentration dependence [8]. In the VB, the calculations find shifting of the valence charge from the N atoms as compared to GaN, which corroborates our identification of the origin of the optical efficiency reduction.

**Resonant x-ray scattering** Resonant SXE spectra with the excitation energies chosen at the two dominant SXA peaks in Fig.1 have shown that the shoulder at the VB bottom scales up and becomes a distinct narrow peak at a binding energy of ~7.4 eV. This is attributed to resonant inelastic soft-X-ray scattering (RIXS) in which coupling of the absorption and emission processes leads to momentum conservation (see, e.g., a review [2]). Our $GaAs_{1-x}N_x$ random alloy can be reconciled with description in terms of wavevectors **k** using a spectral decomposition over the Bloch waves of unperturbed GaAs [1]. The first SXA peak is due to the $t_2(L_{1c})$ state, whose decomposition is dominated by the $L_1$-point. The VB bottom, by analogy with zinc-blende GaN, is dominated by the $L_1$-point as well. The RIXS process couples these states, blowing up the SXE signal in the VB bottom.

**Conclusion** The local electronic structure of N atoms in a diluted $GaAs_{1-x}N_x$ ($x$=3%) alloy, in view of applications in optoelectronics, is determined for the first time using SXE/SXA. Deviations from crystalline GaN, in particular in the CB region, are dramatic. Employing the orbital character and elemental specificity of the SXE/SXA spectroscopies, we identify a charge transfer from the N atoms at the VBM, reducing the overlap between the CBM and VBM wavefunctions, as the main factor limiting the optical efficiency of $GaAs_{1-x}N_x$ alloys. Moreover, a **k**-conserving process of resonant inelastic x-ray scattering involving the $L_1$ derived valence and conduction states is discovered.

*Acknowledgements* We thank L. Gridneva for skillful technical assistance, S. Butorin for sharing his expertise in the SXE data processing, and G. Cirlin for valuable discussions. The work at the Ioffe institute is supported by the NATO Science for Peace Program (SfP-972484) and Russian Foundation for Basic Research (project 02-02-17677).

### References

[1] P.R.C. KENT, A. ZUNGER, Phys. Rev. Lett. **86** (2001) 2613; Phys. Rev. B **64** (2001) 115208.
[2] A. KOTANI, S. SHIN, Rev. Mod. Phys. **73** (2001) 203.
[3] A.YU. EGOROV *et al*, J. of Cryst. Growth **227-228** (2001) 545.
[4] A.M. MINTAIROV *et al*, Phys. Rev. Lett. **87** (2001) 277401.
[5] J. NORDGREN *et al*, Rev. Sci. Instrum. **60** (1989) 1690.
[6] K. LAWNICZAK-JABLONSKA *et al*, Phys. Rev. B **61** (2000) 16623.
[7] C.B. STAGARESCU *et al*, Phys. Rev. B **54** (1996) R17335.
[8] T. MATTILA, S.-H. WEI, A. ZUNGER, Phys. Rev. B **60** (1999) R11245.